\documentclass[sigconf]{acmart}

\AtBeginDocument{%
  }

\setcopyright{acmlicensed}
\copyrightyear{2025}
\acmYear{2025}

\acmConference[SIGIR '25]{the 48th International ACM SIGIR Conference on Research and Development in Information Retrieval}{July 13-18, 2025}{Padova, Italy}

\usepackage{adjustbox}
\usepackage{booktabs}
\usepackage{multirow}

\begin{document}

\title{Beyond Reproducibility: Advancing Zero-shot LLM Reranking Efficiency with Setwise Insertion}

\author{Jakub Podolak}
\affiliation{%
  \institution{University of Amsterdam}
  \city{Amsterdam}
  \country{The Netherlands}}
\email{jakub.podolak@student.uva.nl}
\author{Leon Perić}
\affiliation{%
  \institution{University of Amsterdam}
  \city{Amsterdam}
  \country{The Netherlands}}
\email{leon.peric@student.uva.nl}
\author{Mina Janićijević}
\affiliation{%
  \institution{University of Amsterdam}
  \city{Amsterdam}
  \country{The Netherlands}}
\email{mina.janicijevic@student.uva.nl}
\author{Roxana Petcu}
\affiliation{%
  \institution{University of Amsterdam}
  \city{Amsterdam}
  \country{The Netherlands}}
\email{r.m.petcu@uva.nl}



\begin{abstract}
This study presents a comprehensive reproducibility analysis and extension of the Setwise prompting method for zero-shot ranking with Large Language Models (LLMs), as proposed by \citet{Zhuang_2024}. We evaluate the method's effectiveness and efficiency compared to traditional Pointwise, Pairwise, and Listwise approaches in document ranking tasks. Our reproduction confirms the findings of Zhuang et al., highlighting the trade-offs between computational efficiency and ranking effectiveness in Setwise methods.
Building on these insights, we introduce \emph{Setwise Insertion}, a novel approach that leverages the initial document ranking as prior knowledge, reducing unnecessary comparisons and uncertainty by prioritizing candidates more likely to improve the ranking results. Experimental results across multiple LLM architectures - Flan-T5, Vicuna, and Llama - show that Setwise Insertion yields a 31\% reduction in query time, a 23\% reduction in model inferences, and a slight improvement in reranking effectiveness compared to the original Setwise method. These findings highlight the practical advantage of incorporating prior ranking knowledge into Setwise prompting for efficient and accurate zero-shot document reranking.
\end{abstract}


\maketitle

\section{Introduction}
Large Language Models (LLMs) have rapidly gained popularity in various natural language processing tasks, such as machine translation, question answering, and document ranking \cite{naveed2023comprehensive}. In this study, we specifically focus on their applicability in document ranking. Known for their zero-shot generalization capabilities, LLMs can effectively rank documents without fine-tuning, relying solely on carefully crafted prompts \cite{liu2023pre}. Three key ranking strategies—Pointwise, Pairwise, and Listwise—have been widely explored in the Learning-to-Rank (LTR) literature \cite{liu2009learning}. While these methods differ in both effectiveness and computational efficiency, their performance in zero-shot ranking with LLMs remains underexplored. In particular, there is a lack of comprehensive studies examining how these approaches trade off between effectiveness and efficiency in this setting.

\citeauthor{Zhuang_2024} compare these methods and propose a new ranking method - Setwise - which seeks to balance effectiveness and efficiency by leveraging group-wise comparisons rather than individual document pairs (Pairwise) or full-ranked lists (Listwise). Unlike Pairwise approaches that require $\mathcal{O}(n^2)$ comparisons, Setwise reduces the number of necessary inferences by comparing small sets of documents at once, making it more efficient. Furthermore, unlike Listwise approaches that often struggle with the limited context window of LLMs, the Setwise method mitigates this issue by processing manageable document subsets in each ranking step. The authors apply all methods on retrieving and sorting the top $k$ most relevant documents ($k << n$) from an initially ranked sequence of $n$ documents. They identified the limited context of LLMs, i.e., the limited number of documents that fit in a prompt, as one of the main challenges of LTR with LLMs.

The main contributions of \citet{Zhuang_2024} are:

\begin{itemize}
    \item Analyzing the trade-offs between Pointwise, Pairwise, and Listwise ranking approaches in terms of effectiveness and computational efficiency.
    \item Proposing the Setwise method, which improves efficiency by ranking small groups of documents simultaneously rather than evaluating document pairs independently. This reduces the number of LLM calls compared to Pairwise methods while maintaining strong ranking effectiveness.
    \item Extending Setwise to leverage logit-based ranking within a Listwise setup, enabling a more efficient and interpretable ranking process.
\end{itemize}

Motivated by this foundational work, our study aims to assess the reproducibility and extensibility of the findings presented in the study of \citet{Zhuang_2024}. By replicating key experiments and proposing novel extensions, we seek to validate the robustness of the Setwise method while exploring potential improvements. Specifically, our study brings the following contributions:

\begin{itemize}
    \item Reproducing results for the TREC dataset using Flan-T5 models and other open-source decoder-only models to verify the original findings of \citet{Zhuang_2024}.
    \item Introducing a novel Setwise insertion sort reranking method that utilizes the initial ranking order for higher efficiency, resulting in a 31\% reduction in query time while maintaining strong ranking effectiveness.
    \item Comparing methods proposed by \citeauthor{Zhuang_2024} with our Setwise method, both using modern small LLMs.
    \item Assessing the impact of leveraging initial ranking order on our proposed Setwise insertion method.
\end{itemize}

This reproducibility study not only validates the claims of \citet{Zhuang_2024}, but also expands their work with a new and more efficient reranking method that reduces the number of necessary comparisons and helps with epistemic uncertainty \cite{baan2023uncertaintynaturallanguagegeneration}.

\section{Background}

We begin by introducing the concept of learning-to-rank (LTR). The objective of LTR is to learn a scoring function \( f: \mathbb{R}^n \rightarrow \mathbb{R} \), where the input is a feature vector, and the output is a relevance score \cite{liu-2011}. In the existing literature, three primary methods are used to learn the scoring function \( f \): Pointwise, Listwise, and Pairwise. We define these methods in the context of LTR and discuss their relevance to LLM zero-shot ranking.

\subsection{Pointwise}

The Pointwise approach begins by creating a single feature vector from pairs of queries and documents \cite{liu-2011}. This vector is then passed through the scoring function, which determines the document's relevance. The relevance score can be either continuous or categorical \cite{Li2011ASI}. The function \( f \) is learned by optimizing appropriate loss functions, which may be based on classification or regression tasks \cite{liu-2011, burges2005ranknet, zhu2023large}. Pointwise can be implemented efficiently, with a time complexity of \( O(n) \), where \( n \) is the number of documents for a given query \cite{Zhuang_2024}. However, Pointwise has significant drawbacks. The primary issue is that ranking is not fundamentally a classification or regression problem; it involves predicting the relative relationships between documents rather than their individual relevance. As a result, Pointwise does not generate this relative ordering and does not directly optimize for ranking metrics, such as NDCG@k \cite{liu-2011}.

To apply Pointwise in LLM zero-shot ranking, two methods have been proposed: generation and likelihood. In the generation method, the model is prompted with a question asking whether a document is relevant to the query. This process is repeated for each document in the query. The final ranking is determined by examining the logits or the tokens "yes" and "no" to establish the relative order of the documents. In the likelihood method, the LLM is prompted to generate a query for each document, and the probability of the generated query is used to rank the documents. It is important to note that both methods require access to the logits of the tokens to compute the score, making Pointwise unsuitable for closed-source models, such as GPT-4.

\subsection{Pairwise}

The Pairwise method addresses the limitations of Pointwise by considering the relative order of document pairs. Pairwise takes two documents as input and outputs a preference, typically denoted as \( +1 \) or \( -1 \). Formally, if document \( d_i \) is more relevant than \( d_j \) (i.e., \( y_i > y_j \)), the scoring function should satisfy \( f(x_i) > f(x_j) \). The goal is to minimize the number of inversions required to achieve the optimal ranking, typically defined as: 
\begin{equation}
    \mathbb{L}_{\text{Pairwise}} = \theta(f(x_i) - f(x_j))
\end{equation}

where the function \( \theta \) may vary, such as hinge, exponential, or logistic loss \cite{liu-2011}. While Pairwise improves over Pointwise, it has its own disadvantages. The most significant issue is its time complexity, which is \( O(n^2) \) \cite{ailon2007efficientreductionrankingclassification} since every document must be compared with every other document to determine the optimal ranking \cite{liu-2011}.

\subsubsection{Pairwise Sorted}

The Pairwise Sorted method enhances the basic Pairwise approach by introducing sorting, which leverages data structures to efficiently and selectively determine which documents should be compared. This significantly improves efficiency, as the model does not need to compare two documents that are both clearly less relevant than the current top-\( k \) documents. The method utilizes bubble sort and heap sort, as proposed in previous works \cite{qin2023large, Zhuang_2024}.

When applying Pairwise Sorted to zero-shot LLM ranking, the model is given pairs of documents along with the query and is asked to determine the relevance preference. The time complexity improves with sorting algorithms, resulting in \( O(k \cdot \log_2(N)) \) for heapsort and \( O(k \cdot N) \) for bubblesort, where \( N \) is the number of documents for a query, and \( k \) is the number of top-\( k \) documents we want to retrieve.

\subsection{Listwise}

Listwise approaches take a set of documents for a given query \( q \) and produce an ordered list of these documents as output. This ordered list can be directly compared to the ideal ranking for evaluation. Loss functions for Listwise methods can directly optimize ranking metrics, such as NDCG@k, making them more useful for tasks where improvements in the loss function translate to measurable gains in performance. However, since NDCG is not continuous and differentiable, surrogate metrics such as changes in NDCG must be used. An example of this approach is LambdaRank \cite{burges2010ranknet}.

Implementing Listwise in zero-shot LLM ranking requires some adjustments. For example, the Flan model family used by \citeauthor{Zhuang_2024} has a maximum input length of 512 tokens, which prevents passing all the documents in a single pass. To overcome this, they employed a sliding window of size 4 and made multiple passes over the document list to generate the final ranking \cite{Zhuang_2024, liu-2011}. This results in a time complexity of \( O(r \cdot (n / s)) \), where \( r \) is the number of passes, \( n \) is the total number of documents, and \( s \) is the step size of the sliding window \cite{Zhuang_2024}.

\subsection{Setwise}

Setwise prompting, introduced by Zhuang et al., is designed to address the inefficiencies of Pairwise and Listwise ranking methods when applied to zero-shot document ranking with LLMs. Traditional Pairwise methods require $\mathcal{O}(n^2)$ comparisons, making them computationally expensive, while Listwise approaches often struggle with the limited context window of LLMs, which restricts the number of documents that can be processed simultaneously.

The core innovation of Setwise is that it enables ranking by evaluating small groups of documents (i.e., "sets") in a single LLM inference. Instead of comparing each document pair independently (Pairwise) or attempting to rank an entire list (Listwise), Setwise processes overlapping subsets of documents, allowing for more efficient ranking decisions. This design significantly reduces the number of LLM calls while maintaining competitive ranking effectiveness. Setwise captures relative relevance relationships more effectively than Pairwise approaches, as it considers contextual interactions among multiple documents at once in each inference. Additionally, by limiting the number of documents per inference, Setwise circumvents the context-length constraints that hinder Listwise ranking. The result is a method that balances efficiency and effectiveness, demonstrating strong performance in resource-constrained scenarios.

Zhuang et al. implement Setwise within ranking pipelines using sorting algorithms such as Heapsort and Bubblesort, adapting them to work with set-based comparisons instead of pairwise swaps. This integration of Setwise into classic sorting algorithms further enhances its efficiency, enabling faster and more scalable document ranking. Empirical results from Zhuang et al. show that Setwise outperforms Pairwise and Listwise methods in terms of efficiency while achieving comparable or superior ranking effectiveness. These findings highlight the potential of Setwise as a practical and scalable approach for zero-shot LLM-based document reranking.

\subsection{Insertion Sort Background}

The insertion sort \cite{10.5555/1614191} algorithm works by maintaining an ordered sequence of elements $S$ and an ordered sequence of candidate elements $C$. The algorithm processes each candidate $c_i \in C$ and inserts it at the correct position in $S$ to keep $S$ sorted.

The classic insertion sort algorithm has a worst-case time complexity of $O(n^2)$ as for each of the $n$ candidate elements, there is a possibility it needs to be inserted at the correct position in $S$, which may take up to $n$ operations. More precisely, the time complexity is $O(n + I)$ \cite{10.5555/1614191} where $I$ is the number of inversions: pairs of elements $c_i, c_j$ where $i < j, c_i > c_j$. For nearly sorted arrays $I$ is close to 0, making the algorithm  efficient and closer to $n$ operations. This observation allows us to exploit the fact that the sequence is initially ranked, achieving higher efficiency. In the next subsection, we suggest further optimizations that result in a proposed Setwise Insertion Reranking method.




\section{Methodology}

\subsection{Setwise approach and sorting algorithms}
Original work by Zhuang et al. explores efficient ways to retrieve and sort top $k$ documents from an initially ranked sequence \cite{Zhuang_2024}. Their approach leverages a technique called \textit{Setwise comparison}, where an LLM is used to compare multiple documents simultaneously instead of the traditional pairwise comparisons. To sort (re-rank) an array, we need not only a comparison method but also a sorting algorithm that orchestrates these comparisons in the correct order.

In the classic bubblesort algorithm \cite{10.5555/1614191}, adjacent elements are compared and swapped if necessary - ultimately propagating larger elements toward the end of an array. With setwise comparisons, an element can be evaluated against $c$ neighbors at the same time, allowing it to propagate further in a single pass, thus increasing the algorithm's efficiency.

In contrast, the classical heapsort algorithm builds and maintains a heap \cite{10.5555/1614191}, that is a tree-based data structure in which each parent node is greater (or smaller, depending on the desired order) than its children. The setwise heapsort adapts this method by constructing a max-heap where each node has $c$ children instead of the conventional 2. This adjustment results in a flatter heap structure, which reduces the time complexity of heap updates. To be more precise, assuming that $n$ is the number of the documents to rerank, it requires $O(n + k \log_c n)$ LLM calls - $O(n)$ to construct a heap \cite{10.5555/1614191, Zhuang_2024}, $O(k \log_c n)$ to retrieve top $k$ documents in the correct order.

We noticed that these methods do not utilize the fact that the initial array is already ranked, effectively discarding important signals. We propose two ideas on how to utilize it for even better efficiency, presented in Sections 3.2 and 3.3.

\subsection{Prompts used}

In their work, the authors employ four distinct prompt templates to implement different retrieval methods. For \textbf{Pointwise}, they use the following prompt provided by \citeauthor{sachan-etal-2022-improving} \cite{sachan-etal-2022-improving}:
\begin{quote}
\emph{Passage: \{passage\}, Query: \{query\}. \\
Does the passage answer the query? Answer ``Yes'' or ``No''.}
\end{quote}
When switching to a likelihood-based approach (sometimes referred to as the ``generation'' method to obtain logits), the prompt created by \citeauthor{qin2023large} \cite{qin2023large} is used:
\begin{quote}
\emph{Passage: \{passage\}. \\
Please write a question based on this passage.}
\end{quote}
For \textbf{Pairwise} comparisons, they use the prompts suggested by \citeauthor{qin2023large} \cite{qin2023large} as well:
\begin{quote}
\emph{Given a query \{query\}, which of the following passages is more relevant to the query? \{passage\_1\}, \{passage\_2\}. \\ 
Output Passage A or Passage B.}
\end{quote}
Finally, the \textbf{Listwise} strategy uses the following prompt from \citeauthor{sun2023chatgpt} \cite{sun2023chatgpt}:
\begin{quote}
\emph{The following are \{num\} passages, each indicated by number identifier {}[]. \\
I can rank them based on their relevance to query: \{query\} \\
{}[1] \{passage\_1\} \\
{}[2] \{passage\_2\} \\
\ldots \\
The ranking results of the \{num\} passages (only identifiers) is:}
\end{quote}

\subsection{Informing the LLM about the prior order}

\begin{figure}[H]
    \centering
    \includegraphics[width=1\linewidth]{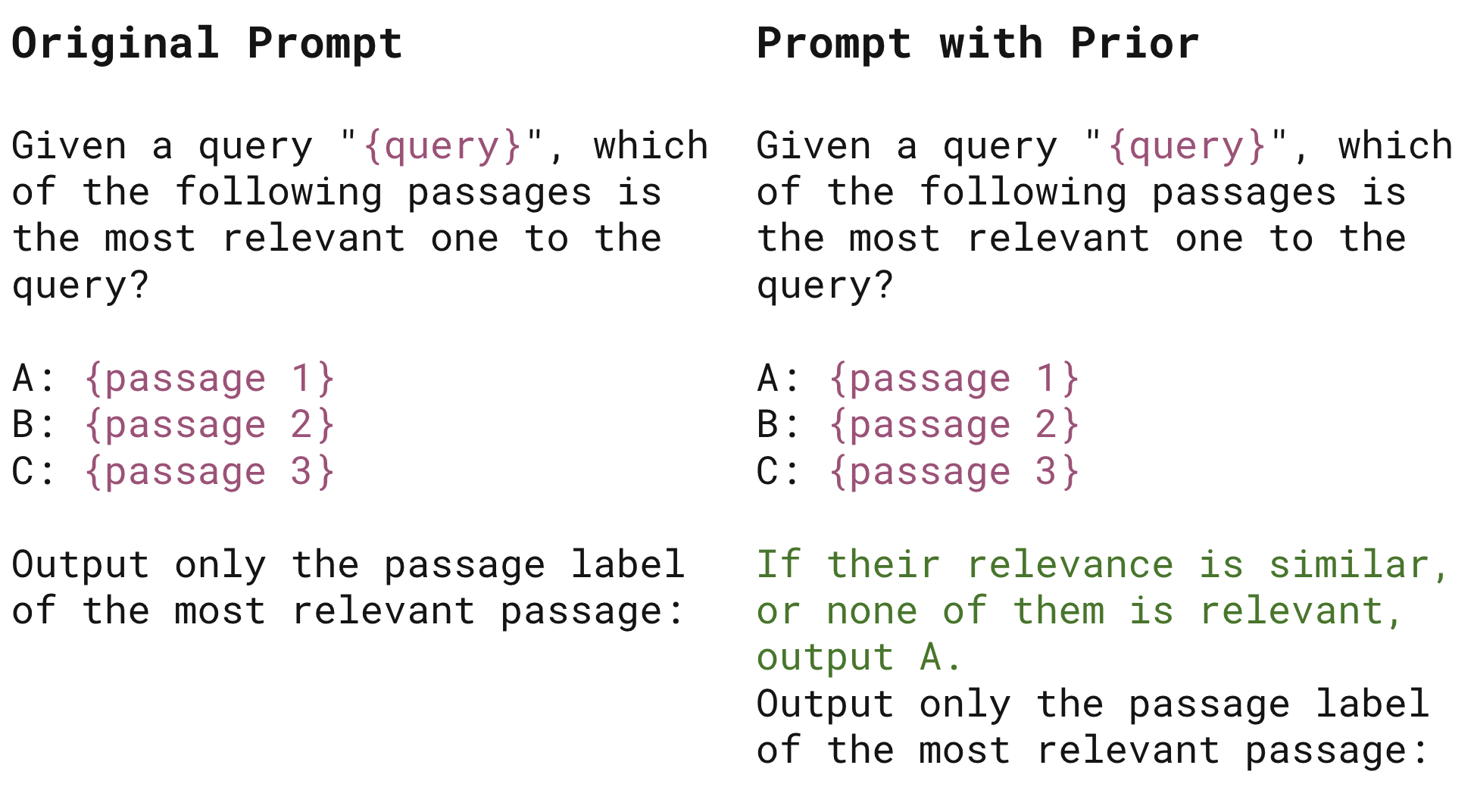}
    \caption{Original Setwise prompt vs our proposed prompt with prior knowledge. We bias the model to return document "A" when uncertain. When constructing a prompt from the template, we put the document with the highest prior (e.g. highest score from BM25) as the document A.}
    \label{fig:prompt-original-prior}
\end{figure}

In the original work by Zhuang et al., one Setwise comparison is realized by (1) creating a prompt with a set of $c$ documents and (2) querying the model to pick the most relevant out of them. We extend the original prompt by asking the model to select the first document when uncertain about the order. The prompt is designed in such a way that the first document is always one for which we have prior information suggesting its relevance. This prior information can include:
a) being ranked higher in the original ranking, or b) being higher in the max-heap. Please refer to (\autoref{fig:prompt-original-prior}) for a comparison of the original and proposed prompt.

\begin{figure}
    \includegraphics[width=1.0\linewidth]{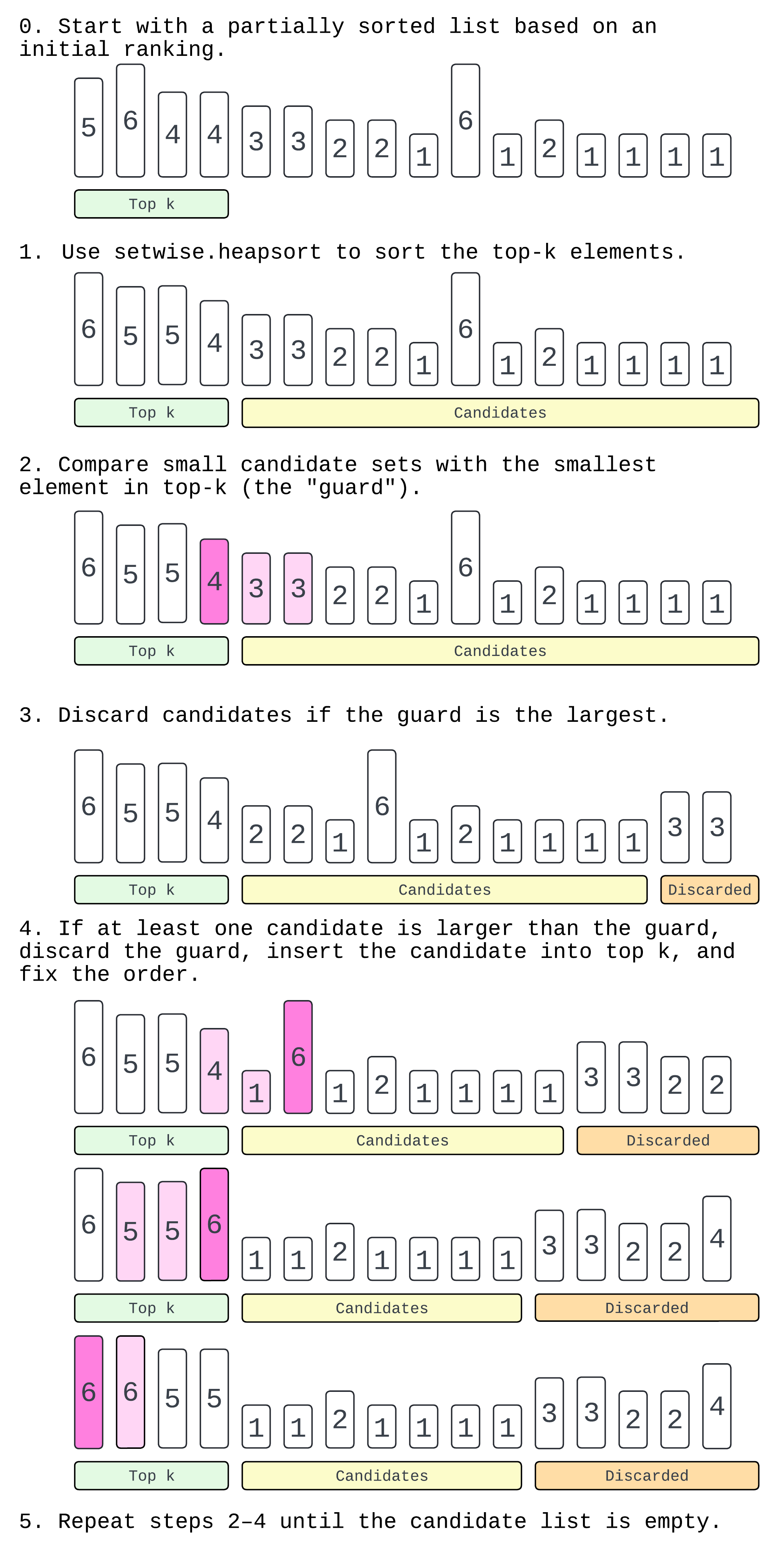}
    \caption{Our proposed Setwise Insertion Sort method for efficient second-stage reranking of top-\textit{k} documents using an LLM. At each step of the algorithm, we maintain only the top-k documents sorted. For each set of candidates, we check if any candidate is larger than the smallest in the top-k. If yes - we promote it to the top-k, and discard it otherwise.}
    \label{fig:insertion_sort}
\end{figure}

We hypothesize that this optimization can lead to improvements in:
\begin{itemize}
    \item \textbf{Efficiency} – by reducing the probability of unnecessary swaps, making the reranking process more stable.
    \item \textbf{Effectiveness} – LLMs can make mistakes and hallucinate. Hallucinations are often correlated with high predictive distribution entropy \cite{xiao2021hallucinationpredictiveuncertaintyconditional} - that is an uncertainty of the model. The literature commonly distinguishes between two types of uncertainty: aleatoric uncertainty, which is related to noise in the training data, and epistemic uncertainty, which stems from insufficient coverage of the representational space during training. This means the model has not seen diverse enough data during training \cite{baan2023uncertaintynaturallanguagegeneration}. While reducing aleatoric uncertainty is difficult, the epistemic can be reduced by adding more diverse data during training. We hypothesize that by biasing the model toward the first document (for which we have a higher prior score), we introduce an additional signal that may help mitigate epistemic uncertainty. Intuitively, when uncertain, the LLM will favor the first document over selecting one at random.
\end{itemize}




\subsection{Setwise Insertion Reranking}
In this subsection, we adapt the classical insertion sort to find and order top $k$ documents by making a forward pass to an LLM. We call the resulting algorithm Setwise Insertion, following the naming convention of Zhuang et al \cite{Zhuang_2024}.

The first observation is that it is possible to keep only the top $k$ documents sorted, to limit the size of the sorted array $S$, thus greatly reducing the time of inserting the candidates $c_i$ at the correct position. At any point of the algorithm, if a candidate document $c_i$ is less relevant than the least relevant document in $S$ and $|S| = k$, the algorithm can discard $c_i$ without performing any further comparisons. To simplify the algorithm even further, we begin it by sorting the top $k$ documents retrieved by the initial ranker (e.g. using Setwise heapsort) and keeping $|S| = k$ fixed throughout the whole reranking process.

The second observation is that we can process the candidates faster using a Setwise approach. When constructing the prompt for the LLM, we choose a set of $c$ documents to compare, where the first one is the least relevant in $S$, and the rest are the candidates. This allows us to find candidates worth adding to $S$ more efficiently. Furthermore, if a candidate $c_i$ should be in $S$, Setwise enables us to find a correct position in $S$ faster, by comparing many documents in $S$ with $c_i$ in one LLM call.

\autoref{fig:insertion_sort} showcases our proposed algorithm. We estimate the time complexity of our Setwise insertion as 
\begin{equation}
    O\Big(k \ log_c \ k + \frac{n}{c} + a \frac{k}{c}\Big)
\end{equation} where $(k \ log_c \ k)$ is the number of comparisons to sort the first $k$ documents using Setwise heapsort,  $(\frac{n}{c})$ to scan the $n$ candidates with sets of size $c$, $a$ equals the number of cases where a candidate is being inserted in $S$, and $(\frac{k}{c})$ is the number of operations to insert the candidate in $S$ where $|S| = k$. For a nearly sorted sequence, $a$ should be small resulting in a high efficiency.

\section{Experimental Setup}

\subsection{Datasets and Models}

For both reproducibility and extension analyses, we use the TREC 2019 and TREC 2020 datasets \cite{trec2019} \cite{trec2020}, while excluding the BEIR datasets due to time and computational constraints. To reproduce the original results, we used the following models: Flan-T5-large, Flan-T5-XL, Flan-T5-XXL \cite{flant5}. We also use LLaMA2-Chat-7B \cite{llama2}, and Vicuna-13B \cite{vicuna}. GPT-3.5 was excluded from our experiments as it is not open-sourced \cite{brown2020language}. 

For the extension analysis, we incorporated Flan-T5-large, Flan-T5-XXL \cite{chung2022scalinginstructionfinetunedlanguagemodels}. Additionally we use LLaMA-3.1-8B-Instruct, LLaMA-3.2-3B-Instruct \cite{dubey2024llama}, and Gemma2-IT \cite{team2024gemma}, to explore further and validate our proposed methods. We decided to omit the Flan-T5-XL model for the extension experiments, because of the limited computation available. 

Our experiments were run on an Intel Xeon CPU with 18 cores, a NVIDIA A100 GPU and 120GB of RAM. This differs from the authors' setup and therefore the latency measures cannot be directly compared.  We made our code available online. \footnote{https://github.com/LeonPeric/llm-rankers}

\begin{figure}
    \includegraphics[width=1.0\columnwidth, trim=3 85 0 0, clip]{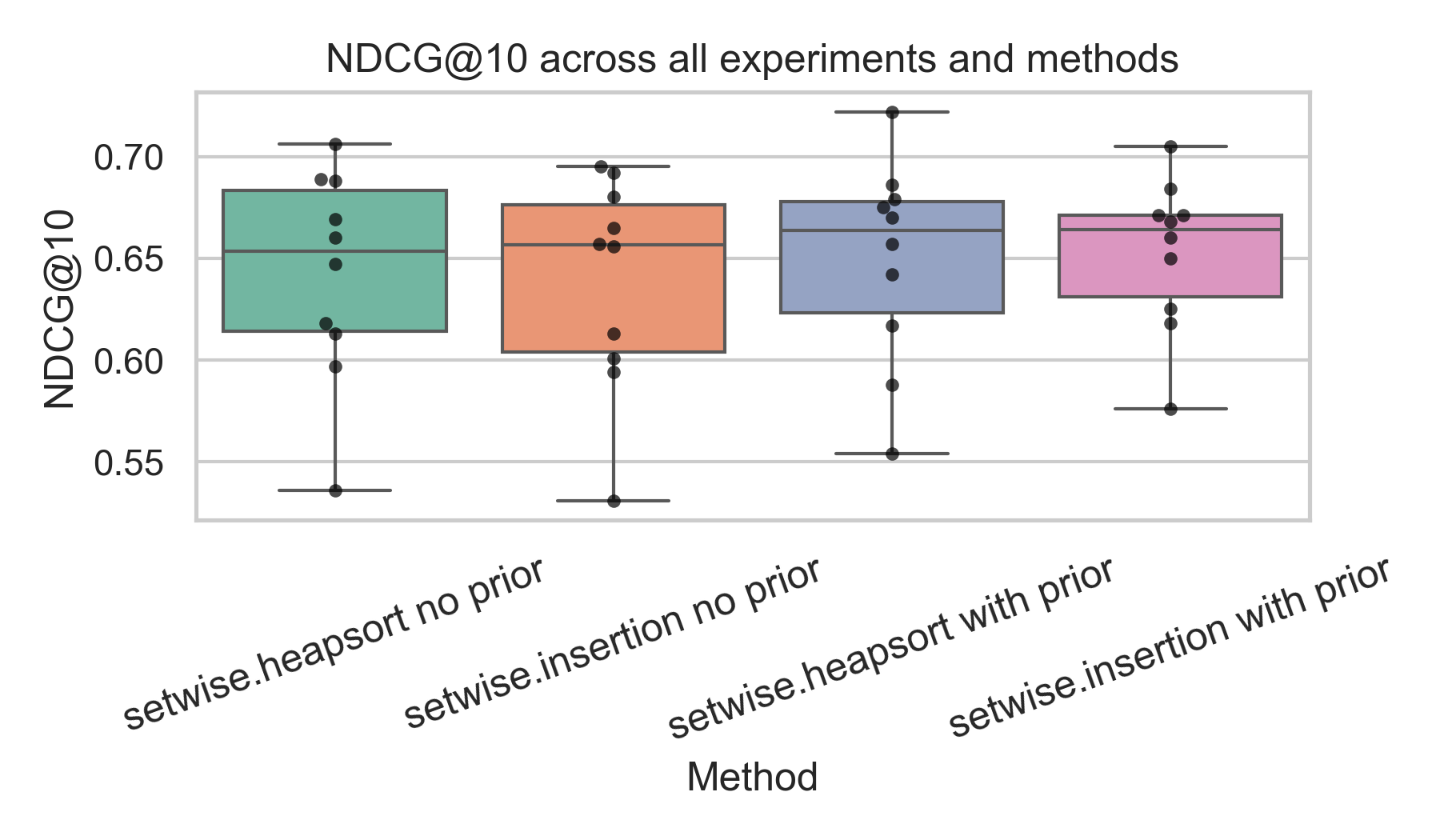}
    \includegraphics[width=1.0\columnwidth, trim=0 26 0 0, clip]{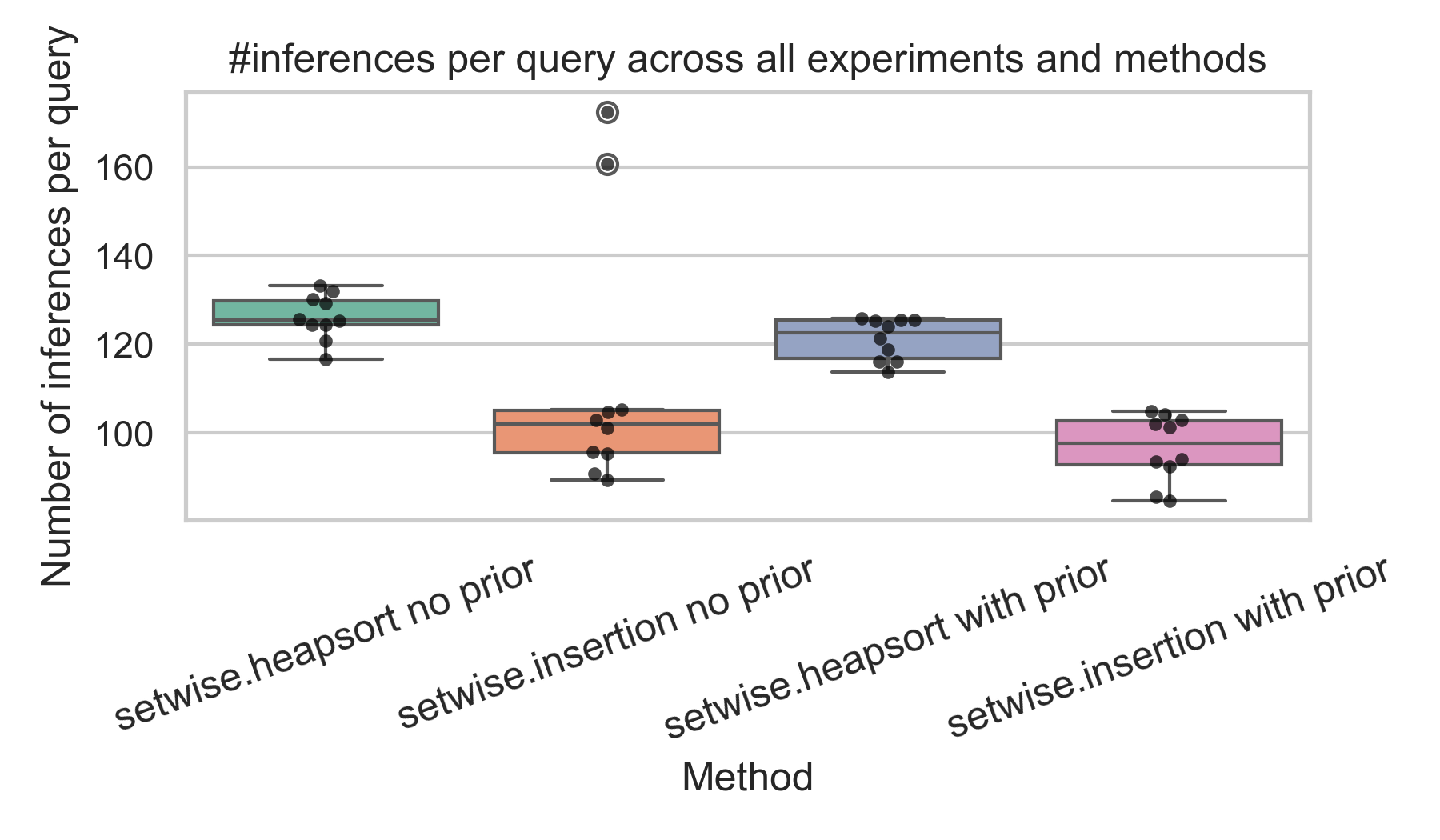}
    \vspace{-0.6cm}
    \caption{NDCG@10 and the number of inferences per query after introducing our two proposed optimizations - insertion sort, and initial ranking prior. Results for all tested models and datasets.}
    \label{fig:boxplot-ndcg}
\end{figure}

\subsection{Setwise Insertion implementation}

When processing candidates in Setwise Insertion, we query the LLM to select the most relevant document from a set of candidates included in the prompt. We explore two approaches to using the LLM for this task: \textit{max compare} and \textit{sort compare}.

\begin{itemize}
\item \textit{Max compare}: We directly use the LLM’s final prediction, i.e., the single most relevant document. If the most relevant document is one of the candidates - we promote it to $S$. In such a case, the other candidates are not discarded, as they may still be worth adding to $S$ in the next iterations.
\item \textit{Sort compare}: Instead of relying on the final prediction, we extract the logits for all candidates in the prompt and sort them by their logit probabilities. This allows us to safely discard all candidates with a lower probability than $s$, improving efficiency. However, this approach requires access to model logits, which is often unavailable.
\end{itemize}
For our extension, we compare the authors' best method (setwise.heapsort) with ours. We repeat the experiments for both their method and ours three times to conduct a more reliable analysis. For the Flan models, we experiment with both \textit{max compare} and \textit{sort compare}. For open-source decoder-only transformers, we use only "max compare," as our setup does not allow access to logits. Unless stated otherwise, Setwise Insertion in this work refers to "Setwise Insertion with prior order" and \textit{sort compare} for Flan models, and \textit{max compare} for other decoder-only models.

\section{Results}

\subsection{Reproducibility Results}

\subsubsection{Encoder-Decoder LLM - Flan}

Table \ref{flan} presents our reproduction results, which replicate the findings from Table 2 of the original study. Our results confirm the observations made by \citeauthor{Zhuang_2024}. They report that all zero-shot ranking methods outperform the baseline BM25, a trend that we also observe in our experiments. Additionally, we notice a similar decline in NDCG as the size increases when comparing Pointwise QLM, consistent with the findings of \citeauthor{Zhuang_2024} 

We also observe a significant improvement when using the Setwise prompt for Listwise likelihood compared to Listwise generation. Therefore, our results closely align with those of \citeauthor{Zhuang_2024}, particularly in terms of NDCG@10 scores. However, there are some discrepancies in the average number of LLM calls. For Pointwise QLM, the yes/no method, and Pairwise all pair, the call counts do not exactly match the original table. This difference arises from our use of varying batch sizes to speed up computation without overloading resources.

Although the average number of prompt and output tokens differs slightly, the differences are not substantial. Furthermore, the average time per query cannot be directly compared to the original table, as we employed a different GPU. However, we can compare the ratios of average times between different methods, and these ratios show comparable results. The variation in GPU architecture and design contributes to the observed differences in absolute times. Lastly, we did not run the Pairwise all pair method on the TREC 2020 dataset due to time constraints.

\begin{table*}[htbp]
\caption{Results for TREC 2019 and TREC 2020.}
\label{flan}
\begin{adjustbox}{width=\textwidth}
\begin{tabular}{lrrrrrrrrrr}
\toprule
\multirow{2}{*}{Method} & \multicolumn{5}{c}{TREC 2019} & \multicolumn{5}{c}{TREC 2020} \\
\cmidrule(lr){2-6} \cmidrule(lr){7-11}
 & \multicolumn{1}{c}{NDCG@10} & \multicolumn{1}{c}{Avg Comp.} & \multicolumn{1}{c}{Avg Prompt} & \multicolumn{1}{c}{Out Token} & \multicolumn{1}{c}{Avg Time} & \multicolumn{1}{c}{NDCG@10} & \multicolumn{1}{c}{Avg Comp.} & \multicolumn{1}{c}{Avg Prompt} & \multicolumn{1}{c}{Out Token} & \multicolumn{1}{c}{Avg Time} \\
\midrule

\multicolumn{11}{l}{\textbf{Large}} \\
BM25 & 0.5058 & -- & -- & -- & -- & 0.4796 & -- & -- & -- & -- \\
Pointwise QLM & 0.5553 & 4.0 & 15,115.63 & 0.0 & 2.90 & 0.5653 & 4.0 & 15,103.28 & 0.0 & 2.15 \\
Pointwise Yes/No & 0.6544 & 4.0 & 16,015.63 & 0.0 & 2.93 & 0.6148 & 4.0 & 16,003.28 & 0.0 & 2.15 \\
Listwise Generation & 0.5612 & 245.0 & 119,126.19 & 2,584.05 & 73.21 & 0.5468 & 245.0 & 119,736.36 & 2,479.05 & 70.49 \\
Listwise Likelihood & 0.6650 & 245.0 & 94,183.21 & 0.0 & 13.36 & \textbf{0.6259} & 245.0 & 95,623.88 & 0.0 & 13.60 \\
Pairwise All-Pair & 0.6660 & 4,950.0 & 2,247,148.05 & 49,500.0 & 460.51 & -- & -- & -- & -- & -- \\
Pairwise Heapsort & 0.6565 & 231.02 & 105,265.77 & 2,310.23 & 20.88 & 0.6189 & 226.76 & 104,893.63 & 2,267.60 & 20.57 \\
Pairwise Bubblesort & 0.6355 & 844.44 & 381,545.49 & 8,444.42 & 75.22 & 0.5866 & 781.55 & 360,961.06 & 7,815.50 & 70.97 \\
Setwise Heapsort & 0.6691 & 125.30 & 40,449.58 & 626.51 & 10.57 & 0.6177 & 124.34 & 40,713.19 & 621.70 & 10.49 \\
Setwise Bubblesort & \textbf{0.6782} & 460.37 & 147,751.02 & 2,301.86 & 38.17 & 0.6229 & 456.23 & 149,433.13 & 2,281.13 & 38.35 \\
\midrule

\multicolumn{11}{l}{\textbf{XL}} \\
Pointwise QLM & 0.5410 & 4.0 & 15,115.63 & 0.0 & 2.68 & 0.5422 & 4.0 & 15,103.28 & 0.0 & 2.69 \\
Pointwise Yes/No & 0.6362 & 4.0 & 16,003.28 & 0.0 & 2.76 & 0.6362 & 4.0 & 16,003.28 & 0.0 & 2.76 \\
Listwise Generation & 0.5684 & 245.0 & 119,174.74 & 2,911.35 & 89.24 & 0.5457 & 245.0 & 119,827.31 & 2,829.30 & 89.22 \\
Listwise Likelihood & 0.6746 & 245.0 & 94,446.95 & 0.0 & 13.42 & 0.6746 & 245.0 & 95,756.95 & 0.0 & 13.24 \\
Pairwise Heapsort & \textbf{0.6917} & 241.77 & 110,089.91 & 2,417.67 & 24.15 & \textbf{0.6917} & 245.56 & 112,989.48 & 2,455.55 & 25.36 \\
Pairwise Bubblesort & 0.6619 & 886.91 & 400,364.74 & 8,869.07 & 87.98 & 0.6619 & 869.74 & 400,499.13 & 8,697.35 & 90.78 \\
Setwise Heapsort & 0.6787 & 129.56 & 41,696.47 & 647.79 & 11.66 & 0.6787 & 128.60 & 42,195.70 & 642.98 & 12.32 \\
Setwise Bubblesort & 0.6755 & 466.91 & 149,902.51 & 2,334.53 & 41.89 & 0.6755 & 464.90 & 152,450.43 & 2,324.50 & 43.46 \\
\midrule

\multicolumn{11}{l}{\textbf{XXL}} \\
Pointwise QLM & 0.5066 & 4.0 & 15,115.63 & 0.0 & 4.46 & 0.4901 & 4.0 & 15,103.28 & 0.0 & 4.47 \\
Pointwise Yes/No & 0.6433 & 4.0 & 16,015.63 & 0.0 & 4.63 & 0.6325 & 4.0 & 16,003.28 & 0.0 & 4.62 \\
Listwise Generation & 0.6604 & 245.0 & 119,319.28 & 2,818.81 & 105.14 & 0.6369 & 245.0 & 119,884.85 & 2,706.53 & 101.15 \\
Listwise Likelihood & 0.7016 & 245.0 & 94,555.19 & 0.0 & 27.21 & 0.6891 & 245.0 & 95,946.98 & 0.0 & 27.53 \\
Pairwise Heapsort & 0.7076 & 239.40 & 109,403.21 & 2,393.95 & 38.75 & \textbf{0.6980} & 241.33 & 111,521.14 & 2,413.30 & 40.08 \\
Pairwise Bubblesort & 0.6787 & 870.0 & 394,205.86 & 8,700.00 & 140.99 & 0.6815 & 855.38 & 396,400.90 & 8,553.80 & 139.47 \\
Setwise Heapsort & 0.7061 & 130.09 & 42,074.74 & 650.47 & 17.59 & 0.6882 & 129.30 & 42,444.33 & 646.48 & 17.66 \\
Setwise Bubblesort & \textbf{0.7124} & 468.47 & 150,780.60 & 2,342.33 & 62.20 & 0.6862 & 466.31 & 153,568.52 & 2,331.53 & 64.89 \\
\bottomrule
\end{tabular}
\end{adjustbox}
\label{tab:flan-reproduction}
\vspace{1cm}
\end{table*}

\subsubsection{Decoder-only LLMs}

Table \ref{reproducibility-decoder-only} presents our reproduction of the experiments from Table 3 in the original paper. We excluded GPT-3.5 from our experiments, as it is not open-sourced, but followed the same setup using the TREC DL 2019 and TREC DL 2020 datasets. Effectiveness was evaluated using NDCG@10, while efficiency was measured by average time (s) per query.

Our results confirm the findings of the original paper: Setwise methods consistently achieve the best overall accuracy for both Vicuna and LLaMA models. This highlights the robustness and effectiveness of Setwise prompting in decoder-only architectures, particularly in balancing efficiency and ranking performance.

\begin{table*}[ht]
\centering
\caption{Performance comparison on TREC DL 2019 and TREC DL 2020 datasets.}
\label{reproducibility-decoder-only}
\begin{tabular}{cccccc}
\toprule
 & \multirow{2}{*}{Methods} & \multicolumn{2}{c}{TREC DL 2019} & \multicolumn{2}{c}{TREC DL 2020} \\
\cmidrule(lr){3-4} \cmidrule(lr){5-6}
 & & NDCG@10 & Avg Time (s) & NDCG@10 & Avg Time (s) \\
\midrule
\multirow{5}{*}{} & \textbf{llama2-chat-7b} & & & & \\
 & listwise.generation & 0.5051 & 132.61 & 0.4762 & 128.45 \\
 & pairwise.bubblesort & 0.5404 & 30.89 & 0.5047 & 27.62 \\
 & pairwise.heapsort & 0.4760 & 19.66 & 0.4434 & 8.30 \\
 & setwise.bubblesort & 0.5902 & 18.42 & \textbf{0.5432} & 18.21 \\
 & setwise.heapsort & \textbf{0.5844} & 9.16 & 0.5410 & 5.12 \\
\midrule
\multirow{5}{*}{} & \textbf{vicuna-13b} & & & & \\
 & listwise.generation & \textbf{0.6511} & 154.09 & \textbf{0.6173} & 152.14 \\
 & pairwise.bubblesort & 0.6219 & 77.13 & 0.5914 & 78.02 \\
 & pairwise.heapsort & 0.6276 & 22.04 & 0.5880 & 21.08 \\
 & setwise.bubblesort & 0.6294 & 28.63 & 0.6115 & 30.16 \\
 & setwise.heapsort & 0.6476 & 8.28 & 0.6008 & 8.28 \\
\bottomrule
\end{tabular}
\vspace{1cm}
\end{table*}

\subsection{The Extension Results}

We compared our improvements against the original Setwise heapsort (without prior knowledge) method \cite{Zhuang_2024}, which was identified by the authors as the best-performing approach. Surpassing this method implies outperforming all other approaches presented in \autoref{tab:flan-reproduction}. The full comparison results are shown in \autoref{fig:insertion-sort-table-full}.

\begin{table*}[ht]
\small
\caption{All results for extension experiments. The methods referred to as "Setwise Heapsort" and "Setwise Insertion" in the paper are bolded. For these two methods, we performed 3 repetitions and presented the mean and 95\% confidence intervals.}
\label{extension_results}
\begin{tabular}{ll|lll|lll}
\toprule
 & \multirow{2}{*}{Methods} & \multicolumn{3}{c}{TREC DL 2019} & \multicolumn{3}{c}{TREC DL 2020} \\
\cmidrule(lr){3-5} \cmidrule(lr){6-8}
 & & NDCG@10 & \#Inferences & Avg Time (s) & NDCG@10 & \#Inferences & Latency (s) \\
\midrule
\multirow{7}{*}{} & \textbf{flan-t5-large} & & & & & \\
\multirow{7}{*}{} & \textbf{Setwise Heapsort no prior} & 0.669 $\pm$ 0.0 & 125.30 $\pm$ 0.003 & 10.54 $\pm$ 0.171 & 0.618 $\pm$ 0.0 & 124.34 $\pm$ 0.0 & 10.54 $\pm$ 0.145 \\
\multirow{7}{*}{} & Setwise Heapsort prior & 0.657 & 116.07 & 9.76 & 0.617 & 116.16 & 9.79 \\
\multirow{7}{*}{} & Setwise Insertion max compare no prior & 0.661 & 98.00 & 8.40 & 0.603 & 100.46 & 8.45 \\
\multirow{7}{*}{} & Setwise Insertion max compare prior & 0.643 & 82.02 & 6.99 & 0.608 & 80.90 & 6.88 \\
\multirow{7}{*}{} & Setwise Insertion sort compare no prior & 0.665 & 95.44 & 5.84 & 0.613 & 95.66 & 5.82 \\
\multirow{7}{*}{} & \textbf{Setwise Insertion sort compare prior} & \textbf{0.671} $\pm$ 0.0 & 92.56 $\pm$ 0.0 & \textbf{5.57} $\pm$ 0.053 & \textbf{0.625} $\pm$ 0.0 & 94.04 $\pm$ 0.0 & \textbf{5.60} $\pm$ 0.219 \\
\midrule
\multirow{7}{*}{} & \textbf{flan-t5-xxl} & & & & & \\
\multirow{7}{*}{} & \textbf{Setwise Heapsort no prior} & \textbf{0.706} $\pm$ 0.0 & 130.09 $\pm$ 0.004 & 17.61 $\pm$ 0.048 & 0.688 $\pm$ 0.0 & 129.30 $\pm$ 0.007 & 17.68 $\pm$ 0.433 \\
\multirow{7}{*}{} & Setwise Heapsort prior & 0.686 & 125.47 & 17.35 & 0.675 & 125.49 & 17.66 \\
\multirow{7}{*}{} & Setwise Insertion max compare no prior & 0.687 & 113.44 & 15.33 & 0.675 & 113.42 & 15.80 \\
\multirow{7}{*}{} & Setwise Insertion max compare prior & 0.685 & 99.00 & 13.79 & 0.672 & 98.97 & 13.75 \\
\multirow{7}{*}{} & Setwise Insertion sort compare no prior & 0.680 & 104.81 & 11.00 & \textbf{0.692} & 105.32 & 10.99 \\
\multirow{7}{*}{} & \textbf{Setwise Insertion sort compare prior} & 0.684 $\pm$ 0.0 & 104.88 $\pm$ 0.001 & \textbf{10.98} $\pm$ 0.036 & 0.671 $\pm$ 0.0 & 104.26 $\pm$ 0.0 & \textbf{10.98} $\pm$ 0.024 \\
\midrule
\multirow{7}{*}{} & \textbf{Llama-3.2-3B-Instruct} & & & & & \\
\multirow{7}{*}{} & \textbf{Setwise Heapsort no prior} & 0.613 $\pm$ 0.0 & 133.30 $\pm$ 0.0 & 4.02 $\pm$ 0.013 & 0.536 $\pm$ 0.0 & 132.01 $\pm$ 0.136 & 3.98 $\pm$ 0.075 \\
\multirow{7}{*}{} & Setwise Heapsort with prior & 0.642 & 118.86 & 3.66 & 0.554 & 113.80 & 3.45 \\
\multirow{7}{*}{} & Setwise Insertion max compare no prior & 0.594 & 172.44 & 5.24 & 0.531 & 160.61 & 4.86 \\
\multirow{7}{*}{} & \textbf{Setwise Insertion max compare prior} & \textbf{0.653} $\pm$ 0.007 & 101.36 $\pm$ 0.067 & \textbf{3.11} $\pm$ 0.093 & \textbf{0.577} $\pm$ 0.004 & 93.56 $\pm$ 0.122 & \textbf{2.86} $\pm$ 0.051 \\
\midrule
\multirow{7}{*}{} & \textbf{Llama-3.1-8B-Instruct} & & & & & \\
\multirow{7}{*}{} & \textbf{Setwise Heapsort no prior} & 0.660 $\pm$ 0.0 & 125.74 $\pm$ 0.006 & 5.51 $\pm$ 0.163 & 0.597 $\pm$ 0.0 & 124.41 $\pm$ 0.0 & 5.56 $\pm$ 0.427 \\
\multirow{7}{*}{} & Setwise Heapsort with prior & \textbf{0.679} & 124.12 & 5.65 & 0.588 & 121.26 & 5.99 \\
\multirow{7}{*}{} & Setwise Insertion max compare no prior & 0.657 & 90.86 & 3.99 & 0.601 & 89.45 & 4.04 \\
\multirow{7}{*}{} & \textbf{Setwise Insertion max compare prior} & 0.668 $\pm$ 0.0 & 85.72 $\pm$ 0.001 & \textbf{3.83} $\pm$ 0.048 & \textbf{0.618} $\pm$ 0.0 & 84.80 $\pm$ 0.0 & \textbf{3.84} $\pm$ 0.144 \\
\midrule
\multirow{7}{*}{} & \textbf{gemma-2-9b-it} & & & & & \\
\multirow{7}{*}{} & \textbf{Setwise Heapsort no prior} & 0.689 $\pm$ 0.0 & 116.63 $\pm$ 0.0 & 9.15 $\pm$ 0.446 & 0.647 $\pm$ 0.0 & 120.77 $\pm$ 0.0 & 9.50 $\pm$ 0.439 \\
\multirow{7}{*}{} & Setwise Heapsort with prior & \textbf{0.722} & 125.84 & 9.85 & \textbf{0.670} & 125.37 & 9.61 \\
\multirow{7}{*}{} & Setwise Insertion max compare no prior & 0.695 & 101.09 & \textbf{7.74} & 0.656 & 103.04 & 8.02 \\
\multirow{7}{*}{} & \textbf{Setwise Insertion max compare prior} & 0.705 $\pm$ 0.0 & 102.95 $\pm$ 0.0 & 8.01 $\pm$ 0.157 & 0.660 $\pm$ 0.0 & 102.02 $\pm$ 0.0 & \textbf{7.88} $\pm$ 0.063 \\
\bottomrule
\end{tabular}
\vspace{0.5cm}
\label{fig:insertion-sort-table-full}
\end{table*}

Our results reveal that Setwise Insertion (no prior) reduces the number of inferences compared to the baseline, as shown in \autoref{fig:boxplot-ndcg}. Introducing prior knowledge of the initial ranking into the LLM prompt slightly improves both the efficiency and effectiveness of the algorithms. When applying two proposed optimizations simultaneously, we observe an \textbf{average 31\% decrease in query time (9.41s to 6.27s), a 23\% reduction in LLM inferences (126.2 to 96.6), and a modest improvement in NDCG@10 (0.642 to 0.653)}. This demonstrates that our method is both more efficient and effective than the previous best approach.

\begin{figure}
    \includegraphics[width=1.0\columnwidth, trim=0 0 0 10, clip]{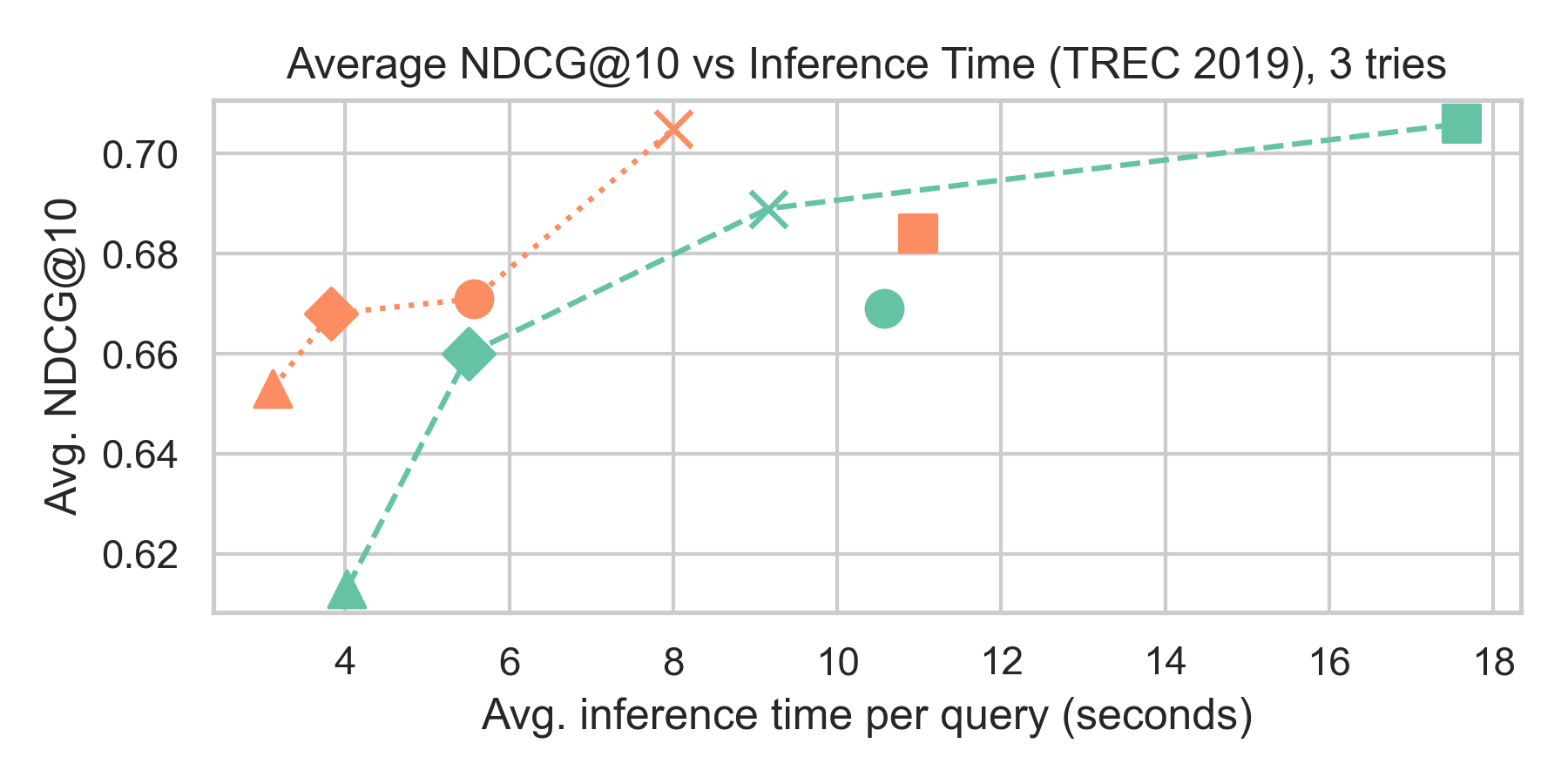}
    \includegraphics[width=1.0\columnwidth, trim=0 0 0 10, clip]{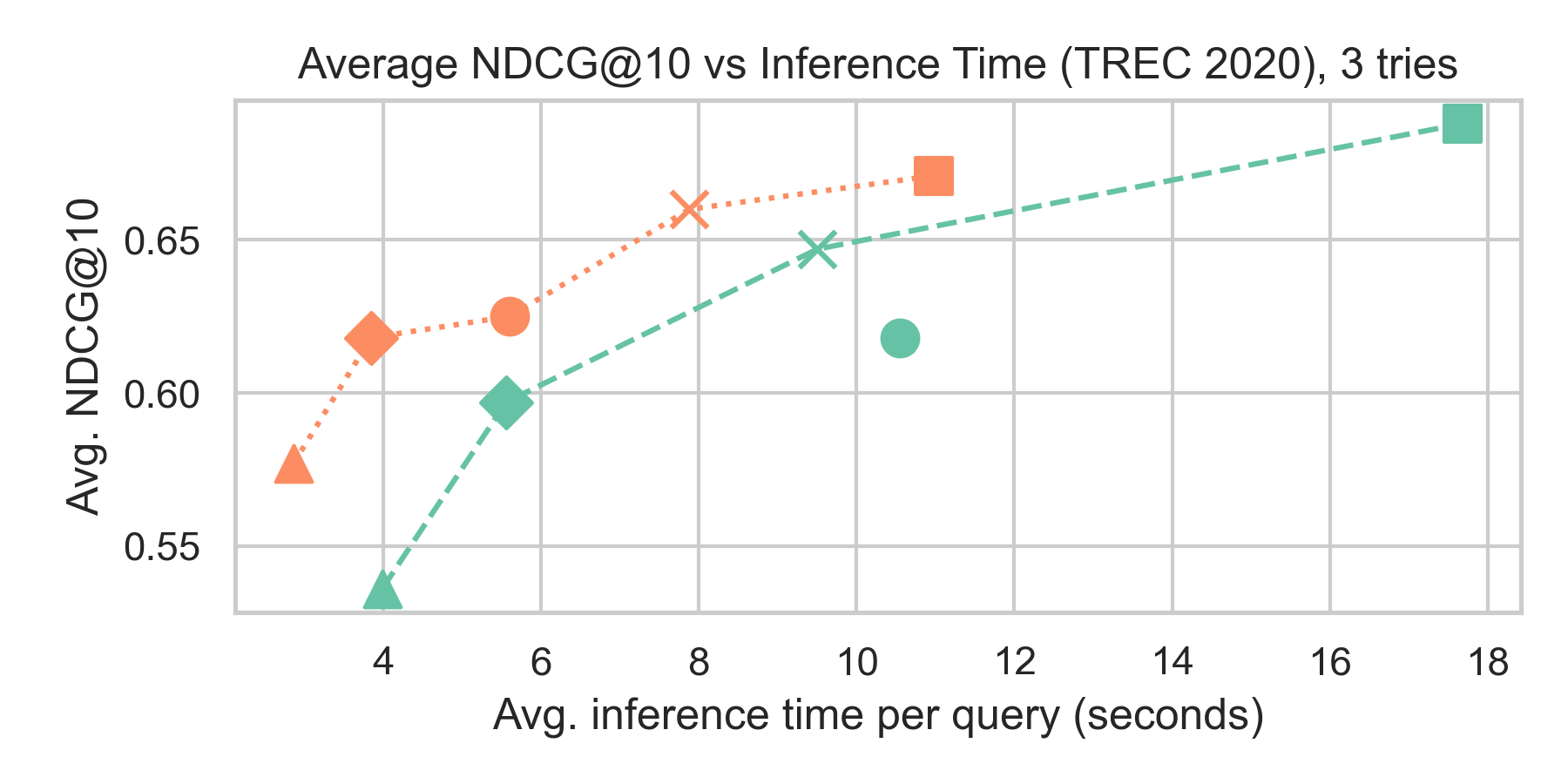}
    \includegraphics[width=1.0\columnwidth, trim=0 0 0 13, clip]{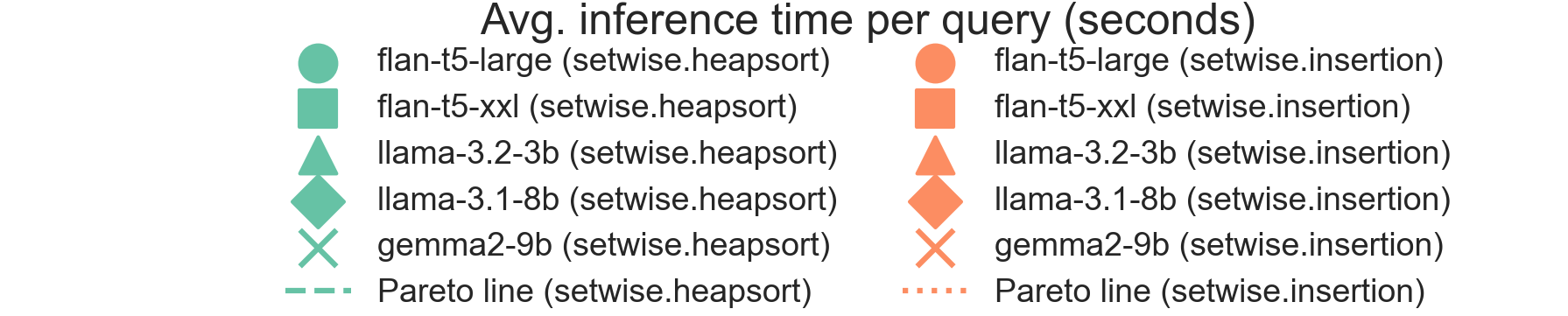}
    \caption{Average NDCG@10 (3 runs) for setwise.heapsort (no prior) and setwise.insertion (with prior) across all tested models. The original author's method is in green (dashed line), and ours is in orange (dotted line). Separate results for TREC 2019 and TREC 2020.}
    \label{fig:pareto-ndcg-models}
\end{figure}

\autoref{fig:pareto-ndcg-models} shows the 3-run average NDCG@10 and inference times for two methods, across five models and two datasets. Dashed lines represent the Pareto frontier for each method and dataset, which highlights the optimal trade-off between efficiency and effectiveness. Beyond this frontier, improving one metric would require sacrificing the other. Our Setwise Insertion method consistently outperforms Setwise Heapsort across both datasets, except for the Flan-t5-xxl model, where introducing Setwise Insertion results in a slight drop in effectiveness.

These results demonstrate that our optimizations to Setwise reranking—both insertion sort and the inclusion of prior ranking knowledge—significantly enhance both efficiency and effectiveness, leading to state-of-the-art performance in LLM top-k reranking.

\section{Discussion}

Our reproducibility study confirms the validity of the results presented by Zhuang et al., affirming that their Setwise reranking method is both effective and efficient for zero-shot document ranking tasks using large language models (LLMs).

We validated the claims of \citeauthor{Zhuang_2024} regarding the efficiency and effectiveness of Setwise methods. Using open-source models, such as Flan-T5, Vicuna and Llama, we reproduced the results on the TREC 2019 and 2020 datasets. Our findings confirmed that Setwise prompting outperforms Pointwise, Pairwise, and Listwise approaches, striking an optimal balance between computational efficiency and ranking performance.

Building on this foundation, we proposed the Setwise Insertion method, which leverages prior knowledge from initial document rankings. By incorporating this information into the reranking process, our approach reduces unnecessary computations and enhances stability without compromising effectiveness. Experimental results demonstrated significant gains, including a 31\% reduction in query time and a 23\% decrease in LLM inferences, along with slight improvements in NDCG@10 scores compared to the original Setwise Heapsort method. These advancements establish Setwise Insertion as the new state-of-the-art for efficient and effective reranking.

Future work could explore extending these experiments to include datasets such as BEIR, which were part of \citeauthor{Zhuang_2024} but omitted in this study due to computational constraints. Additionally, testing a wider range of models could provide further insights into the generalizability of the proposed methods. Lastly, future research could explore whether Setwise can be further improved through a combination of different or more advanced sorting algorithms.

\subsection{What was easy}

Reproducing the core experiments and integrating our extensions were relatively straightforward due to the well-structured and clean codebase provided by \citeauthor{Zhuang_2024}. The modularity of their implementation facilitated seamless incorporation of our proposed methods, including the Setwise Insertion approach. This ease of use underscores the importance of providing clear and reproducible code in academic research.

\subsection{What was difficult}

Certain aspects of the study posed challenges. Understanding the nuances of the Listwise methods required an in-depth examination of the codebase and experimental setup. The naming conventions for Listwise optimizations were sometimes ambiguous, necessitating additional effort to clarify their meanings. Furthermore, reproducing results involved running a substantial number of experiments, which was computationally intensive and occasionally delayed by server downtimes. These challenges, while manageable, highlight the complexities of working with resource-intensive machine learning experiments.

\subsection{Communication with original authors}
We did not communicate with the original authors and instead resolved all challenges independently.

\begin{acks}
We are grateful to Roxana for her guidance and to the University of Amsterdam for providing computational resources.
\end{acks}

\bibliographystyle{ACM-Reference-Format}
\bibliography{sample-base}


\end{document}